# Middleware-based multi-agent development environment for building and testing distributed intelligent systems

Francisco José Aguayo-Canela[1] · Héctor Alaiz-Moretón[1] · María Teresa García-Ordás[1] · José Alberto Benítez-Andrades[2]  · Carmen Benavides[2] · Paulo Novais[3] · Isaías García-Rodríguez[1]

## Abstract
The spread of the Internet of Things (IoT) is demanding new, powerful architectures for handling the huge amounts of data produced by the IoT devices. In many scenarios, many existing isolated solutions applied to IoT devices use a set of rules to detect, report and mitigate malware activities or threats. This paper describes a development environment that allows the programming and debugging of such rule-based multi-agent solutions. The solution consists of the integration of a rule engine into the agent, the use of a specialized, wrapping agent class with a graphical user interface for programming and testing purposes, and a mechanism for the incremental composition of behaviors. Finally, a set of examples and a comparative study were accomplished to test the suitability and validity of the approach. The JADE multi-agent middleware has been used for the practical implementation of the approach.

✉ José Alberto Benítez-Andrades
jbena@unileon.es

Francisco José Aguayo-Canela
francisco.aguayo@ieee.org

Héctor Alaiz-Moretón
hector.moreton@unileon.es

María Teresa García-Ordás
mgaro@unileon.es

Carmen Benavides
carmen.benavides@unileon.es

Paulo Novais
pjon@di.uminho.pt

Isaías García-Rodríguez
isaias.garcia@unileon.es

[1] SECOMUCI Research Group, Escuela de Ingenierías Industrial e Informática, Universidad de León, Campus de Vegazana s/n, C.P. 24071 León, Spain

[2] SALBIS Research Group, Department of Electric, Systems and Automatics Engineering, University of León, Campus de Vegazana s/n, León, 24071 León, Spain

[3] Algoritmi Centre/Department of Informatics University of Minho, Braga, Portugal

## 1 Introduction

The proliferation of devices with Internet connection capabilities in the so-called Internet of Things (IoT) is a trend that is generating an overwhelming amount of new streams of data. These data are crucial to the operation of the systems where the devices are located but must be properly managed to obtain useful information for decision making. The distributed nature of these systems demands decentralized architectures for the management and control of the IoT devices, including tasks such as monitoring or security assurance. The multi-agent paradigm has proven to be a convenient approach for building this kind of decentralized management and control systems [10]. This "agentification" idea underlies the notion of "building intelligence" on the IoT devices, by including an agent inside each of the IoT devices or by using an agent that "represents" the device.

In any case, the "intelligence" provided by the agents may be based on the use of rules, different flavors of logic and other deliberative mechanisms and Artificial Intelligence (AI) techniques [17]. Many existing cognitive functions used for building intelligence into the agent are based on the reactive model, using an event-driven mechanism that, with the aid of a set of rules, allows the agent to sense its environment and trigger its behavior when a change is detected [7]. This kind of approaches have been applied to several domains as healthcare [18], domestic electricity cost control [1] or dynamic caregiver routing problems [15].

An especially useful application scenario for multi-agent systems is IoT security and privacy [9, 11]. Many of the security-oriented tools existing today rely on the definition of a set of rules for detecting meaningful events or a set of given patterns in the network traffic, as is the case of intrusion detection systems (IDS) or firewalls. Thus, incorporating these rule-based reactive systems into the agents of a multi-agent system seems a convenient approach for implementing distributed security IoT environments [9], but the development of such rule-based multi-agent systems is a difficult, time-consuming task.

There are two main approaches to build rule-based multi-agent systems applied to IoT scenarios. The first approach is based on the integration of a rule language (and engine) into an agent component of an existing multi-agent platform [14] and the second involves buiding agent and multi-agent systems capabilities by programming them into a rule language [19]. While the later may be useful in some IoT scenarios where the agents must run within constrained devices, the first approach takes advantage of the full features of the multi-agent platform. For example, in the case of JADE (Java Agent Development Framework) [5], this platform can be used to build robust solutions including the use of the FIPA-compliant Agent Communication Language (ACL) or built-in functionalities for encryption, authentication and authorization.

This research aims at designing and building a development environment for easing the construction and debugging of multi-agent systems that use rules for implementing the cognitive capacities of the agents. The solution is based on a set of tools and functionalities incorporated into the multi-agent middleware. Here, the term middleware must be understood as the combination of an agent framework (the software tools needed for building the skeleton of a multi-agent system) and an agent platform (the software package providing the functionalities for deploying and running distributed multi-agent applications) as stated in [6].

The set of hypotheses that guide this work relate to the features and measures that the solution will have to exhibit and accomplish, making it more suitable for the application scenario (developing distributed management solutions for the IoT) than existing solutions. These desired features can be stated as follows:

- The integration of the rule engine into the agent must be loosely coupled; in such a way that the solution does not depends on the particular implementation technology of the rule engine.
- As it is expected that this kind of agents are involved in a great number of communication processes, the execution of the reasoning must not block the agent, allowing it to keep receiving communications from other agents in the platform while performing the reasoning.
- The solution must adhere to well-stablished multi-agent platform standards.
- The development of the agents in the platform must be incremental, with aids for testing not only the own agent's knowledge base, but also the cooperative problem solving involving many agents.

This set of hypotheses directs the design and development of the proposed platform. The solution is validated by running some experiments for comparing its performance to another similar approach and testing if the desired features are achieved in the different prototypes to be constructed.

The core components of the proposed framework accomplishing the above hypotheses are:

- The integration of a rule engine into the agent, creating a specialized *rule-based agent* component.
- The implementation of a communication mechanism between *rule-based agents*.
- The creation of a wrapping agent component (working as a *development environment*) that extends the main agent component in the platform by exposing a very rich graphical user interface for building and testing *rule-based agents*.
- The modularization and externalization of the agent behaviors, and the construction of a mechanism for the incremental composition of behaviors into the agent.

The multi-agent middleware chosen for testing the practical solution is JADE. This choice is based on its maturity, the size of the user community, and the use of behavior-based agents, which can be exploited to integrate the rule engine processes as behaviors.

Figure 1 shows a diagram of the proposed middleware architecture. The rule engine (A) used by a given JADE agent is loosely coupled to it through a software interface (B) according to the technology of the engine. The rule-based agent will communicate with other rule-based agents by using the standard ACL (Agent Communication Language) (C) developed by FIPA (Foundation for Intelligent Agents) using a set of technology-neutral concepts for describing actions in the rule engines that are stored in a shared ontology (see Table 1). As well as the agent-to-agent, also the agent-to-rule engine communication is achieved by using FIPA-compliant ACL messages that the agent sends to itself (D). Section 2 describes the integration of the rule engine into the agent and these communication processes.

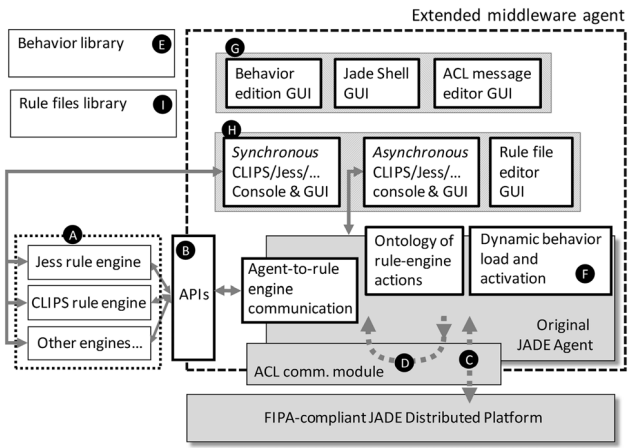

**Fig. 1** Framework for the proposed solution

Besides the integration of the agent and the rule engine, the proposed solution includes a set of facilities to ease the design, development and debugging of rule-based multi-agent systems. The basic JADE Agent class has been extended to build an incremental behavior composition system by using a library of externalized behaviors (E) and a mechanism for the dynamic load and incorporation of these behaviors into the agent (F). This functionality is detailed in Sect. 4.

Some graphical user interface windows have also been built into the extended JADE Agent class to aid the programmer in the process of building the knowledge base for each agent and test the distributed multi-agent system prior to putting it into production. The "agent management tab" (G) includes a behavior edition window, a JADE Shell editor and an ACL message visualization and edition window; all of them are described in Sect. 3.1. On the other hand, the "rule engine management tab" (H) includes a file editor for creating and editing rules that are stored in an external library (I), a synchronous shell for direct interaction with the given rule engine used by the agent and an asynchronous shell where the user can interact both with the rule engine of the same agent or with any rule engine of any agent in the platform. Section 3.2 gives further details about these windows and their functionalities.

The rest of the paper is organized as follows: Section 5 describes the set of experiments used for the validation and testing of the desired features stated in the hypotheses. Some discussion about the results obtained is presented in

**Table 1** Codes and description for possible activities to be communicated between rule-based agents

| Code | Description |
| --- | --- |
| LOAD_FILE | Load the file indicated as a parameter |
| LOAD_FACTS | Loads the facts file indicated as a parameter. |
| LOAD_FROM_RESOURCE | Loads the given resource file indicated as a parameter. |
| LOAD_FROM_STRING | Loads data from a CSV file indicated as a parameter. |
| LOAD_ASSERT_STRING | Loads facts from a string. |
| LOAD_BLOAD | Memory restoring from a bin file. |
| LOAD_SLOAD | Memory restoring from a plain text file. |
| RUN_INFINITELY | Run indefinitely up to the end of rule activations. |
| RUN_NUMBER_OF_CYCLES | Run a given number of cycles. |
| RUN_ONCE_THEN_BATCH | Run and give the control back to the Shell. |
| RUN_INNER_SHELL | Execute the internal Shell. |
| MAKE_RESET | Perform a reset command. |
| MAKE_CLEAR | Perform a clear command. |
| MAKE_MEMORY_DUMP | Perform a security backup. |
| MAKE_ASSERT_STRING | Inserts a fact from a string. |
| MAKE_BUILD | Compile a query. |
| EVAL_COMMAND(*) | Evaluate a sentence. |
| SET_INPUT_BUFFER_COUNT | Requests the number of input characters entered. |
| APPEND_INPUT_BUFFER | Appends to the given command. |
| SET_UNWATCH | Not to analyze (debugger). |
| SET_WATCH | Analyze (Facts, Modules, etc.). |
| GET_FACT_SLOT | Get an slot value. |
| FACT_INDEX | Move the cursor in the fact list. |

Sect. 6. Finally, Sect. 7 is devoted to future work and conclusions.

## 2 Integrating a rule engine into an agent: the rule-based agent

The integration of the rule engine into the agent is crucial to achieving the coordination of the agents and also the flexibility needed to build and debug a multi-agent system, the detailed description and discussion of the integration can be found in [2]. The implementation is based on the extension of the basic *Agent* JADE class to create the "*rule-based agent*" class, which is an agent with capacities for integrating and communicating with an associated rule engine.

The objectives of the integration of the rule engine into the agent are:

– Neutrality concerning the particular technology of the rule engine, for example, CLIPS [13], JESS [12] or Drools [20].
– Agents with a rule-based behavior must be able to communicate to other rule-based agents.
– Actions performed in the rule-based system of a given agent must be fully and exclusively controlled by the agent, separately from other activities or behaviors.
– The rule engine associated with an agent must not block the basal agent behavior while performing the reasoning.
– The design must help and ease the development of rule-based multi-agent applications.

The framework proposed in this work uses an uncoupled integration of the agent with its rule engine, in the sense of being independent of the engine technology by using a programming interface mechanism for interaction. The rule engine is exclusively devoted to the agent, which distinguishes this approach from other ones where the reasoning is built as a service in a special agent devoted to solely perform the execution of the rules that other agents demand [4].

### 2.1 Using an interface for managing the rule engine from the agent

The interaction of the agent and its corresponding rule engine is performed by using an interface (called *RBEngine*) that allows the agent to interact with any rule engine technology by creating an according implementation. The initialization method for the agent includes the code for the creation of the *RBEngine*, due to this, the agent has full control over the functionalities and the responses coming from the engine and, there is no need to decide the kind of inference engine technology to be used for the agent until the very moment of its creation in the platform.

The object that represents the rule engine is unique, and is subordinated to the execution thread of the agent. It is created when the agent is incorporated into the multi-agent platform, not as part of the behavior lifecycle of the agent.

### 2.2 Inter-agent communication mechanism

The communication mechanism between agents with integrated rule engines complies with the FIPA specifications, using ACL messages and a domain ontology for storing the valid set of message contents.

The set of actions that a rule-based agent can ask another one to perform on its rule engine is limited to a predefined set of concepts (see Table 1) that represent the usual activities of these kind systems (loading facts and rules, executing a number of firing cycles, query facts and rules, performing a reset or clear command, etc.).

These actions can be invoked by a *rule-based agent* when communicating to another rule-based agent, but also by a human (working within the enriched interface shell of an agent, see Section III.B.3) that wants to communicate with any other *rule-based agent* in the platform for development or testing purposes. These two kinds of interaction are distinguished by a field value within the content of the ACL message, this way, the receiving agent is aware of whether the message comes from another agent or from a human developer.

### 2.3 Communication of the agent with its rule engine

Figure 2 details the mechanism for the communication between an agent and its rule engine when a message arrives from another agent in the platform. The sender agent is represented as A1 in figure 2 (note that it is duplicated for coherence with the time line used for representing the communication process). When a *rule-based agent* (A2 in figure 2) receives an ACL message from A1 containing actions that are to be performed by its rule engine, the agent, initially, captures this message (first loop with duration te in figure 2) and replies to the sender agent the acceptance (or not) of the action (this reply is not represented in figure 2 for clarity). Then, the receiver agent creates and sends itself a new ACL message containing the same action proposed by the sender, as well as the identification for both the conversation and the sender agent. The action is placed in the receiving agent's message queue to be finally processed by its rule engine when it is ready to do so (second loop with duration $t_p$ in Fig. 2).

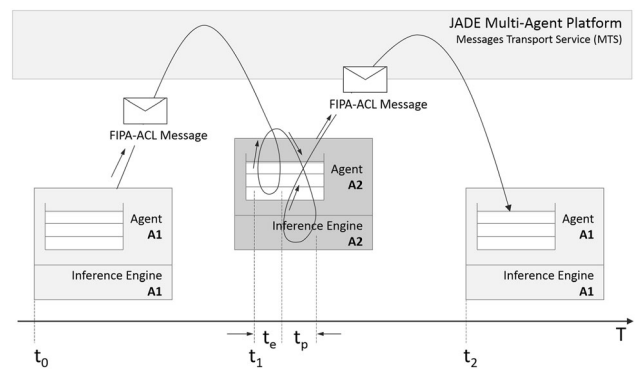

**Fig. 2** Schematic view of an agent A1 sending an ACL message targeted at the rule engine in agent A2

During all this process, the receiving agent is able to keep communicating with other agents from the platform because its functions are not blocked. Once the rule engine processes and executes the indicated set of actions, the sender is informed about the result of these activities (it receives the results at time $t_2$ in Fig. 2). This mechanism is based on a twofold implementation of a FIPA interaction protocol (the first one for the sender agent to destination agent communication and the second for the destination agent to itself – towards its rule engine –), allowing a private, controlled and ordered use of the rule engine of the given agents. Using a FIPA interaction protocol is a coherent decision for the implementation of the communication process, between agents and also between an agent and its rule engine.

### 2.4 Execution of the rule engine activities in a threaded behavior

A finite-state-machine (FSM) behavior is responsible for processing a new message when the rule engine of the agent is ready to do so. The first state of this FSM consists of a listening behavior capturing the message that the agent previously arranged and put in its own message queue. The second state is implemented in an execution behavior, obtaining the contents included in the message, and moving them into the rule engine. A threaded behavior is used to wrap this execution behavior with the objective of not interrupting the execution thread of the agent with the activity of the rule engine. The third state is the response behavior, it starts when the activity of the rule engine ends (and the execution behavior thread no longer exists). This behavior obtains the output provided by the engine, builds the message to be sent as a response and delivers it to the requesting agent including the initial identification for the conversation. Finally, this response behavior returns control of the finite-state-machine to the listening behavior and, if there are any new ACL messages for the rule engine in the message queue of the agent, the next one is processed.

## 3 The development environment

A core component of the solution is the "*development environment*", built as an agent extending the basic *Agent* JADE class. It is primarily intended to ease the development and debugging of agents with rule-based behaviors. When in development and debugging time, the *rule-based agent* must be invoked as a "*development environment*" agent in the platform. This invocation causes the agent to launch with an enriched graphical user interface with a set of functionalities for building and testing *rule-based agents*.

This solution allows the interactive modification of the agent internal code, as well as the interaction with its own rule engine, or even with any other rule engine in any of the agents in the platform. The graphical interface of the *development environment* agent has two main tabs (see Fig. 3). Tab 1 is called the "agent management tab" and tab 2 is the "rule engine management tab". Their functionalities are presented in the next sections.

### 3.1 The agent management tab

The agent management tab includes the following windows:

– The JADE shell window.
– The behavior editing window.
– The message viewing and editing window.

#### 3.1.1 The JADE shell window

The JADE shell window (see Fig. 4) contains the Bean-Shell component [21] connected to the agent to allow a human to program in Java and having direct access to the JADE API objects, the methods and properties of the agent, the methods and properties of the rule engine, etc. It allows to build new classes and to instantiate new objects to be incorporated into the agent, to look at the help files, to perform unitary tests, to watch the message queue, to build new messages and send them, to build and test new behaviors, etc.

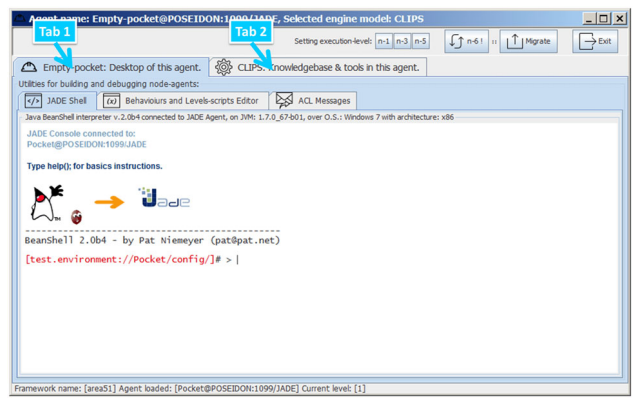

**Fig. 3** The graphical user interface of the development environment

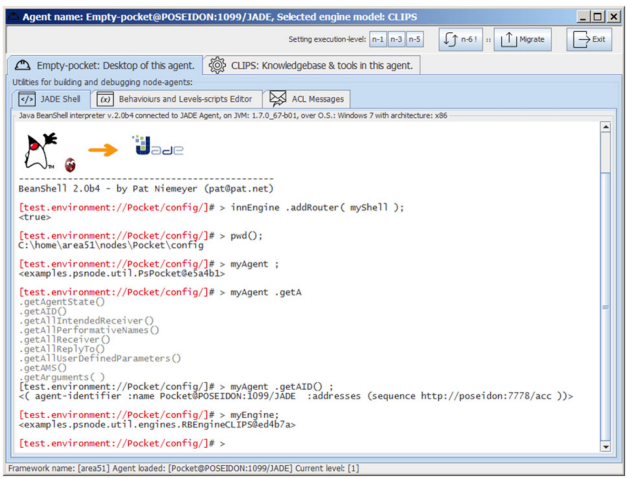

**Fig. 4** JADE shell window

The BeanShell component also has a non-graphical mode, that is used when the agent is in production (without this graphical user interface) in order to read and load the behaviors of the agent at runtime (see Sect. 4).

### 3.1.2 The behavior editing window

The behavior editing (see Fig. 5) contains a text editor making use of the RSyntaxTextArea component (https://bobbylight.github.io/RSyntaxTextArea/). It is used for loading and modifying the behavior files of the agent (see Sect. 4). It has syntax highlighting and word auto-completion capabilities. On the left part of the window, a list shows the set of all the possible behavior files for the agent to be loaded into the editor.

### 3.1.3 Message editor and trace window

The message editor window (see Fig. 6) contains a partial implementation of the *testAgent* component in the *jade.tools.testagent* package distributed with the JADE middleware. It allows to watch the events in the message queue of the agent and manually build new messages.

## 3.2 The rule engine management tab

The "rule engine management tab" includes the functionalities to interact with the associated rule engines of the multi-agent system. It is composed of three windows:

– The file editor, for the expert system managed by the agent.
– The synchronous shell, for communicating with the local rule engine.
– The asynchronous shell, for communicating with any remote rule engine in another agent in the platform.

### 3.2.1 File editor window

Figure 7 shows the file editor window. It is used for editing expert system files locally. It is based on the *RSyntaxTextArea* component and includes CLIPS and Jess syntax highlighting. The files created here can be later loaded into the agent's rule engine or sent remotely to another *rule-based agent*.

### 3.2.2 The synchronous shell window

The synchronous shell allows a human to interact directly with the rule engine of the agent. It performs a direct connection from the graphical interface to the inner rule

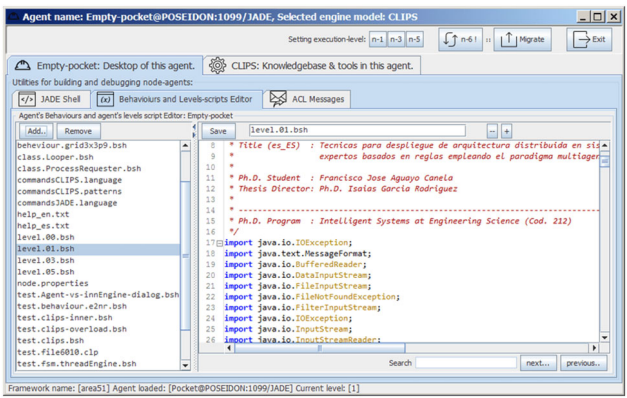

**Fig. 5** Behavior editing window

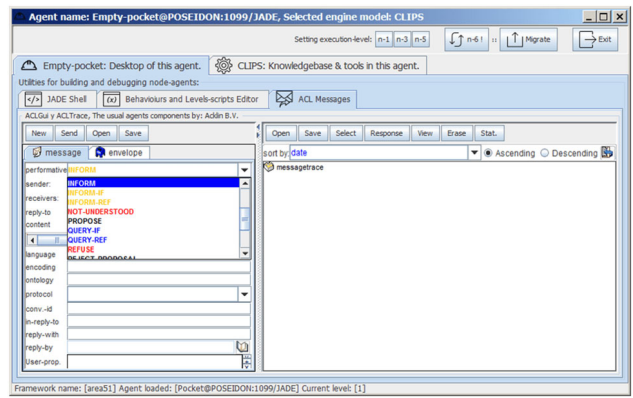

**Fig. 6** Message editing window

🖉 Springer

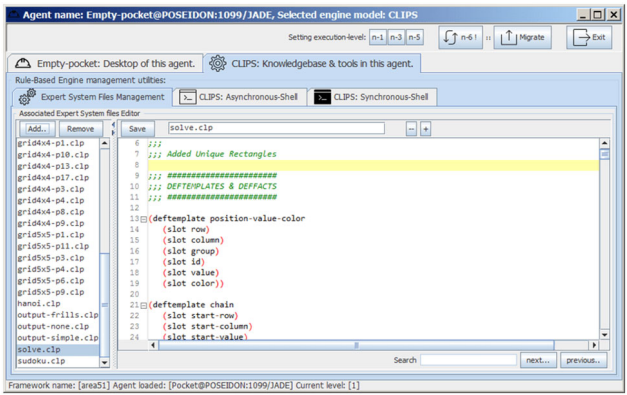

**Fig. 7** Message editing window

engine, emulating a shell of the underlying technology (CLIPS in the case of Fig. 8). This window should only be used during the initial phases of the agent development, and not during execution, where the asynchronous shell is preferred so as not to block the agent operation.

### 3.2.3 The asynchronous shell window

This shell allows the communication with the rule engine of any agent in the platform, including the one of the agents where the interaction takes place (see Fig. 9). The command introduced in this window is included in an ACL message that is sent to the rule engine of the destination agent. It allows the communication with the agents and their rule engines at runtime, using an interaction protocol, without blocking the agent behaviors or the engine execution, as was described in Sect. 2.3.

Once the command is introduced in the upper text area, it can be dispatched by using the combination Shift+Ctrl+Enter, or clicking on the "Excecute!" button. The list on the left allows selecting which of the agents in the platform will be the destination agent, including the own local agent (denoted by the word "itself" in the list). The command is sent to the destination agent in an ACL message and, once the rule engine of that agent processes the instructions, the results are sent back to the sender in an ACL *Notification* message associated with the conversation thread created at the beginning of the interaction. There is a text area at the bottom of the window where the responses of the destination agent (usually the results of the processing of the instructions by the rule engine) will appear. As a result of the solution designed, the window is not blocked while waiting for the response, neither is the destination agent. So, new commands can be sent even to the same destination agent.

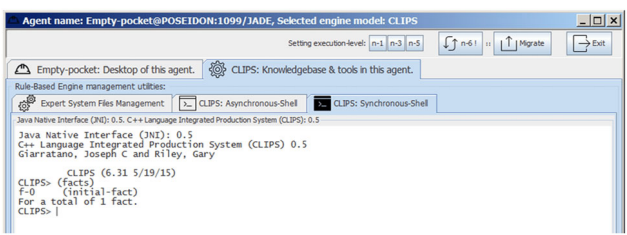

**Fig. 8** The synchronous shell window

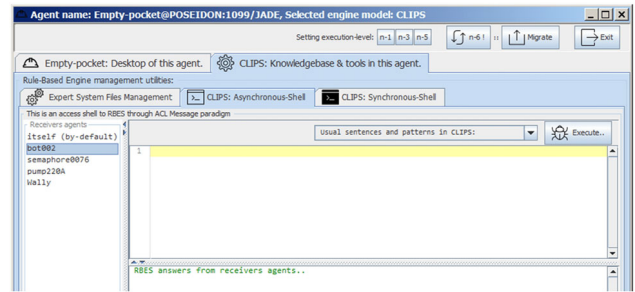

**Fig. 9** The asynchronous shell window

## 4 Externalization and incremental composition of behaviors

The externalization of behaviors allows an agent to load or modify its behaviors by loading and processing them in real-time from local files. The Java interpreter, incorporated into the agents (see Sect. 3.1) is responsible for the processing of these external files and the incorporation of the behaviors in the task manager at a proper time to avoid collisions and incoherences.

This dynamic process of loading the agent behaviors is the base for the incremental composition of behaviors mechanism. The final, complete, behavior of the agent can be composed of different behaviors that can be loaded one at a time. This way, the behavior can be tested step by step, starting with the simplest or basal ones. To ease this modularity and progressivity in the construction of the final behavior, the agents are initialized by going through a series of steps very similar to the "runlevels" found in UNIX-like operating systems.

Five runlevels are defined; each of them has an associated script file associated where the behaviors to be loaded is indicated. Table 2 shows the different runlevels and the associated processes that occur in each of them.

The execution level of the agent can be controlled with the buttons "n-1", "n-3", "n-5" and "n-6!" in the graphical user interface (see Fig. 10).

In practice, these runlevels can be used to incrementally test the functionalities of the agents, for example when building complex behaviors or when testing coordination mechanisms with other agents.

**Table 2** Runlevels and corresponding actions

| Level | Process |
|---|---|
| 0 | The setup() method for the agent finished its execution. The agent is already incorporated into the multi-agent platform and its status is active. The script file [level.00.bsh] is loaded and interpreted. |
| 1 | The script file [level.01.bsh] is loaded and interpreted, which results in the load of the "basal" behaviors for the agent. |
| 3 | Load and interpretation of the script [level.03.bsh]. Activation of behaviors loaded in level [1], objects of the type Behavior that appear in the behavior collection are also loaded. |
| 5 | Load and interpretation of the script [level.05.bsh]. Activation of the behaviors that were loaded in level [3]. Wheever the scripts in [level.05.bsh] are processed, the agent is considered in the state "in service", and the execution level is set to [5]. |
| 6 | The script [level.06.bsh] contains commands that result in a "hot reboot" of the agent, which means that the agent is not removed from the platform, but its active behaviors are stopped and removed from the agent. Following, the execution level [0] is entered. |

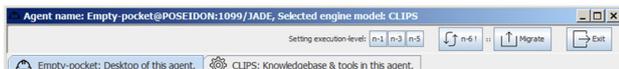

**Fig. 10** Buttons for runlevel activation

## 5 Sample tests

A fully functional, self-contained test environment is available for download at the following URL: https://seco muci.com/research/MAS/IMAS/. Instructions for deployment and use are included, and a tutorial describing different tests to be performed shows the different functionalities of the *development environment* in a practical context.

The *rule-based agents*, in their default configuration, include behaviors devoted to communicating with other agents in the platform and with the rule engines of these agents. They have no specific cognitive functionality. The first test to be accomplished is to launch the JADE platform and to create two *rule-based agents* named "Agent200" and "Agent300". These agents are invoked in "debugging mode" and so will be wrapped by the class that exposes the graphical user interface for development and debugging purposes.

By using the asynchronous shell window, a human can interact with any rule engine of any agent in the platform. For example, Fig. 11 shows how, from Agent300, a (facts) command can be sent to the remote Agent200 rule engine, and how this agent returns the information obtained after issuing this command on its rule engine. The destination agent for the command is chosen in the left list among all the existing agents in the platform, in this case, Agent200 is selected. The bottom part of the window shows the conversation maintained between Agent300 and Agent200. The blue color is used for representing the message from the origin to the destination, while the red color represents the response from the destination. As was stated in Sect. 2.3, the command is sent to Agent200 as an ACL message, and Agent200 is responsible for processing the message, deciding if the command can be passed to its rule engine and building a new ACL message for sending the results back to Agent300.

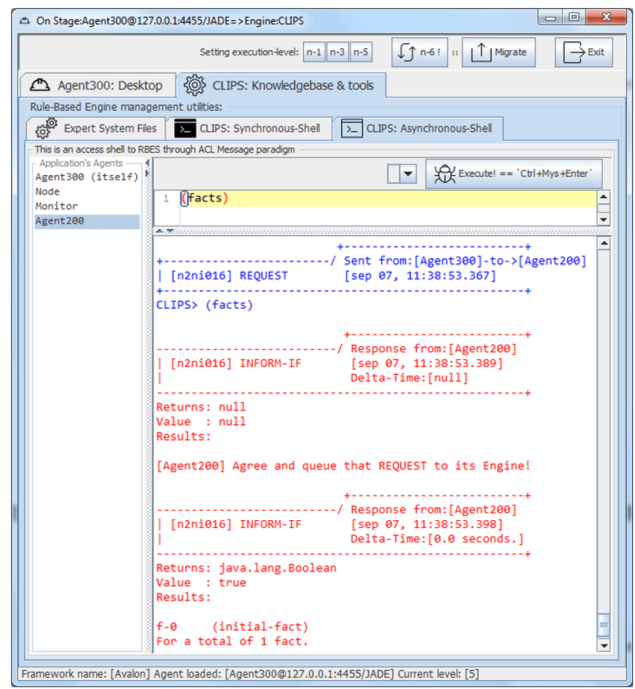

**Fig. 11** Inserting a new fact in a remote rule-based agent

The human developer, by using Agent300, can send facts and rules to Agent200 and ask for their execution. Figure 12 shows how Agent300 sends the command for a fact creation to Agent200 (Fig. 12a), a command for the creation of a rule (Fig. 12b), and finally, the (run) command in order to execute the rule system, showing the results in the response (Fig. 12c).

More complete examples are provided at https://seco muci.com/research/MAS/IMAS/, showing, for instance, how to incorporate a rule-based system as an agent

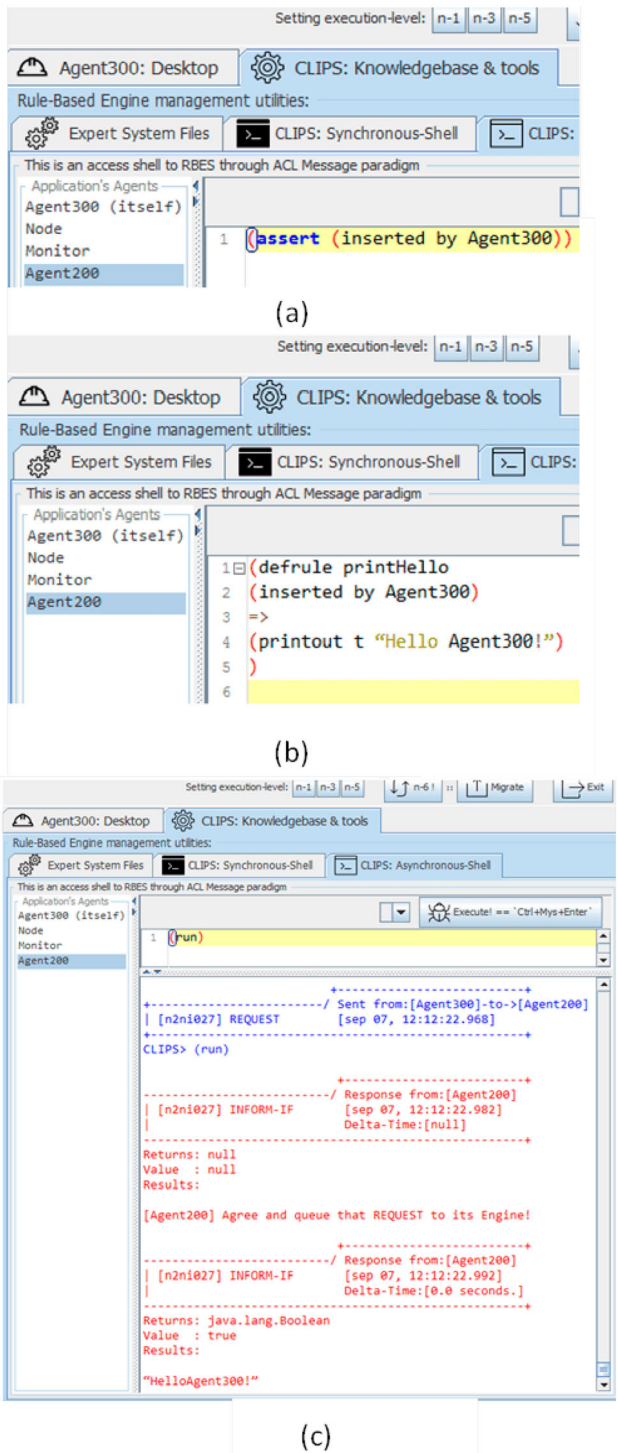

**Fig. 12** Execution of a rule-based system on a remote rule-based agent

behavior and letting other agents send problem data and retrieve the solution found.

### 5.1 Validation of the solution

The set of characteristics described in Sects. 3 and 4 constitutes an enriched middleware framework that no other multi-agent platform solution has. The sample test described in the previous paragraphs demonstrates that the desired set of functionalities established for the middleware in the initial hypotheses were accomplished.

Section 2 describes the integration of the rule engine into the agent, as well as the communication mechanism through which the rule-based agents in the platform can collaborate. To validate and test the proposed approach, a comparative study has been designed and implemented. The study compares the solution described in this paper to the integration described in [8]. The study can be reproduced by following the indications and using the software at https://secomuci.com/research/MAS/IMAS/validation.

The solution proposed in [8] integrates the JESS rule engine into a JADE agent by taking advantage of the shared programming language used by both technologies (Java). This solution uses a new Agent class, *JessAgent*, that can communicate with an instance of the JESS rule engine. It uses a cyclic behavior, along with facilities for reading and loading external rule files. Control of the engine is accomplished by performing callbacks to the *JessSend* function (an internal function declared in JESS) from the agent.

The study compares the performance and response times of an agent from the framework proposed in this paper and a *JessAgent* from the solution proposed in [8], where the initial *JessAgent* class was transformed into a new *HLCjessAgent* class to make it compatible with the JESS version used by the approach proposed in this paper to put the agents on equal conditions. The agent from the framework proposed in this paper is called *DPSNodeAC* during this section. The experiment was carried out using a single PC with an AMD A9-9410 RADEON R5 processor.

The experiment consists in making a third agent, called *Analyzer*, generate a number of messages to be sent to the agents under test (*HLCjessAgent* and *DPSNodeAC*). The messages can be of two types:

– Presence request messages (used for testing if the agent is alive in the platform). The response is a simple acknowledgement for confirming the presence of the agent. The usual response time for this kind of message in the mentioned computer was about 300 ms when the agent is free from other reasoning processes.
– Requests messages asking for solving sudoku problems of different difficulties. The response to these messages is the solution found for the given problem and so they take the agent more time to respond than the presence request messages. There are four different sudoku problems to be solved, with solving times (in the rule base) from around 200 ms to 2500 ms.

The sequence of messages used is shown in Table 3. A total of 40 messages were sequentially generated for each agent. The first four are of type "presence request" (p in Table 3), the fifth is of type "sudoku" (S in Table 3), then nine more "presence request" messages are sent and one "sudoku" follows, this sequence is repeated twice, ending with five more "presence" messages. Each message is scheduled to be sent from the *Analyzing* agent every 250 ms. So, the entire message sequence is generated within 10 seconds.

The *Analyzer* agent is responsible for sending the messages and capturing the corresponding responses, annotating the time at which the message was issued and the time when the corresponding response from the agent arrived, the difference is the corresponding delay for the given message.

Figure 13 shows the delays in the responses for each message for the *HLCjessAgent* (a) and for the *DPSNodeAC* agent (b). As can be seen, the approach described in this paper outperforms the results of the other solution. For the four first messages (of "presence request" type) the delay in the responses is similar for both agents. But, after the first sudoku request message (sudoku request messages are represented as dotted columns in Fig. 13), the *HLCjessAgent* shows delays that are much bigger than those from the *DPSNodeAC* agent for the following presence request messages. Moreover, it can be seen that, each time a sudoku request message is processed, the *HLCjessAgent* is affected with bigger delays for the next presence request messages.

Taking into account the total time, that is, the time from the delivery of the first message from the *Analyzer* agent to the reception of the last response from the corresponding destination agent, the last message from the *DPSNodeAC* agent arrived at 10392 ms, while the last response from

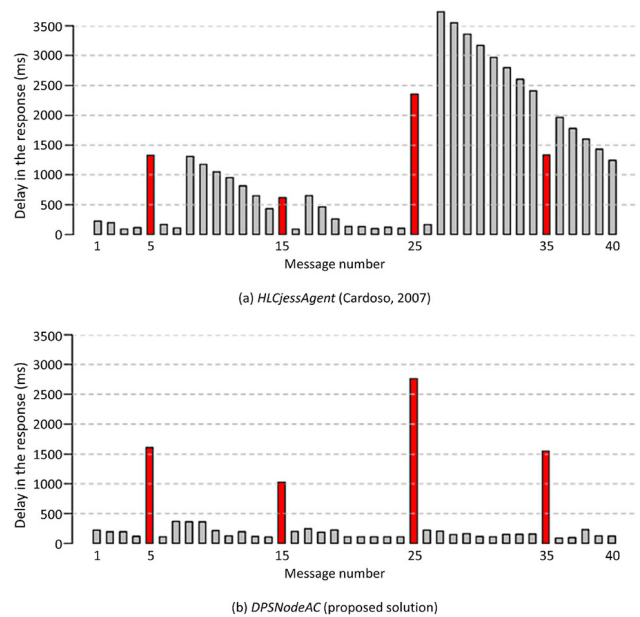

**Fig. 13** Delays in the responses of HLCjessAgent vs DPSNodeAC

*HLCjessAgent* reached the *Analyzer* agent at 12124 ms, that is, exceeding in more than 1700 ms.

## 6 Discussions

This work describes an enriched middleware for multi-agent platforms that can be used as a development framework for building multi-agent systems. The solution is based on the flexible, loosely coupled and technology-independent integration of a rule engine into an agent, as opposed to usual solutions that rely on a less flexible, highly coupled integration [16]. The framework includes the development of specialized agents with graphical user interfaces for easing the development and debugging processes, as well as a mechanism for incremental incorporation of externalized behaviors into the agents. The approach used for implementing this integration maintains a careful separation between the tasks of the agent and those of the rule engine, but each *rule-based agent* has associated its own rule engine, compared to other approaches like the EMERALD framework [4] that uses the "reasoning as a service" paradigm.

Compared to the integration of a rule engine into an agent described in [8], the solution proposed outperforms the results using a simple stress test, showing that the proposed integration allows the agent to keep on receiving and processing messages while the rule engine is working on the solution of a given problem. In a real scenario, this would speed up the whole multi-agent system if these kinds of agents were used.

**Table 3** Sequence of messages for the experimental test of the solution

| Message# | 1 | 2 | 3 | 4 | 5 | 6 | 7 | 8 | 9 | 10 |
|---|---|---|---|---|---|---|---|---|---|---|
| Type | p | p | p | p | S | p | p | p | p | p |
| Message# | 11 | 12 | 13 | 14 | 15 | 16 | 17 | 18 | 19 | 20 |
| Type | p | p | p | p | S | p | p | p | p | p |
| Message# | 21 | 22 | 23 | 24 | 25 | 26 | 27 | 28 | 29 | 30 |
| Type | p | p | p | p | S | p | p | p | p | p |
| Message# | 31 | 32 | 33 | 34 | 35 | 36 | 37 | 38 | 39 | 40 |
| Type | p | p | p | p | S | p | p | p | p | p |

The solution proposed in this paper is inspired by the "edge computing" paradigm [3], where the data processing is achieved locally to where they have been generated; in this case, the local environment of the agent includes its associated rule engine. In a typical IoT scenario, especially those related to monitoring and security solutions, the data processing and communication processes among agents would be very numerous, and so the solution presented in this research seems to be adequate when dealing with the use of such rule-based agents in the edge of IoT infrastructures. The multi-agent based architecture is one of the paradigms usually employed for the implementation of IoT edge-based solutions [22].

The examples in Sect. 5 show the flexibility of the approach for building and debugging complex distributed rule-based systems. Interaction with the human developer was used to show how a developer can use the system, but this kind of interaction is only meant to be used at developing or debugging time. When in production, the agents will be created in the platform as *rule-based agent* objects without these graphical user interfaces, and so only agent-to-agent communication will be possible.

## 7 Conclusions

The framework described in this paper allows and eases the incremental development and debugging of rule-based multi-agent systems. The strategy used for integrating the rule engine into the agent made it possible to obtain a more flexible and faster solution than other similar ones, which is an advantage in knowledge-intensive multi-agent applications. The framework can be downloaded from https://github.com/dpsframework-/dpsFrameworkBuilder/releases.

There is a need for implementing security and trust mechanisms in order the agents can authenticate and decide if the message, including actions involving its rule engine come from an authorized agent. At this moment, the only mechanism that the agent implements is to test whether the message comes from another agent in the platform, but the designed solution eases the implementation of more complex authentication and authorization mechanisms because of the double implementation of the interaction protocol described in Sect. 2.3.

Future work includes the implementation of collaboration and coordination mechanism for the *rule-based agents*, and the use of Semantic Web formalisms (RDF, SWRL, etc.) for representing rules and facts into the agents. The development environment is currently being used for building an intelligent distributed system for IoT security based on the multi-agent paradigm.

**Publisher's Note** Springer Nature remains neutral with regard to jurisdictional claims in published maps and institutional affiliations.

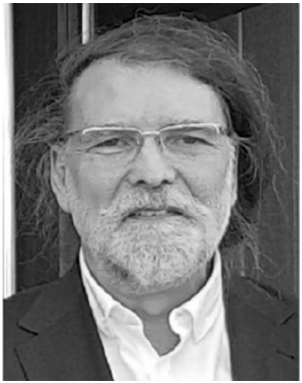

**Francisco José Aguayo-Canela** Ph.D. received the B.Sc. degree in mechanical engineering from the Polytechnic School of the University of Seville, in 1989, and the M.Sc. degree in communication and information technology management from the Faculty of Computer Science of the University of Seville, in 2011, and the Ph.D. degree in Intelligent Systems Applied to Engineering from School of Industrial, Computer and Aerospace Engineering of the Leon University, Spain, in 2017. From 1989 to 2001, he worked at the Regional Office of the Institute of Cartography and Statistics, on the development of macroeconometric models and the analysis of evolution shown by time series of municipal variables. Until 2006, he headed the Information Technology Department of the Regional Econometric and Sociological Studies Foundation. From 2006 to 2011, he has formed key part of the team for the documentary and archaeological catalog project of the Regional Office for the Conservation of Historical Heritage. From 2012 to the present, he has been integrated as head of cybersecurity, at the telecommunications department of the Regional Office. His research interests include: intrusion detection algorithms, statistical modeling, real-time multi-agent systems, and assisting agents.

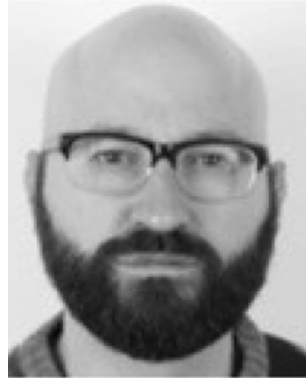

**Héctor Alaiz-Moretón** Ph.D. received his degree in Computer Science, performing the final project at Dublin Institute of Technology, in 2003. He received his Ph.D. in Information Technologies in 2008 (University of Leon). He has worked like a lecturer since 2005 at the School of Engineering at the University of Leon. His research interests include knowledge engineering, machine & deep learning, networks communication and security. He has several works published in international conferences, as well as books and scientific papers in peer review journals. He has been member of scientific committees in conferences. He has headed several Ph.D. Thesis and research projects.

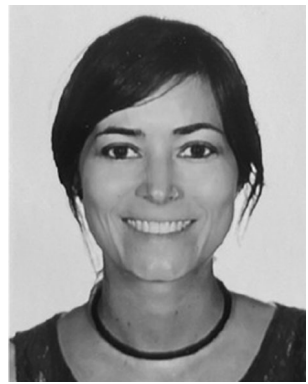

**María Teresa García-Ordás** Ph.D. was born in León, Spain, in 1988. She received her degree in Computer Science from the University of León in 2010, and her Ph.D. in Intelligent Systems in 2017. She was a recipient of a special mention award for the best doctoral thesis on digital transformation by Tecnalia. Since 2019, she works as teaching assistant at the University of León. Her research interests include computer vision and deep learning. She has published several articles in impact journals and patents. She has participated in many conferences all over the world.

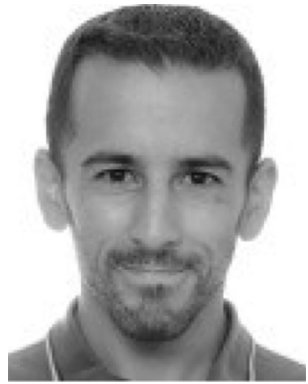

**José Alberto Benítez-Andrades** Ph.D. was born in Granada, Spain, in 1988. He has received his degree in Computer Science from the University of León, and the Ph.D. degree in Production and Computer Engineering in 2017 (University of Leon). He was part time instructor who kept a parallel job from 2013 to 2018 and since 2018 he works as teaching assistant at the University of Leon. His research interests include artificial intelligence, knowledge engineering, semantic technologies. He was a recipient of award to the Best Doctoral Thesis 2018 by Colegio Profesional de Ingenieros en Informática en Castilla y León in 2018.

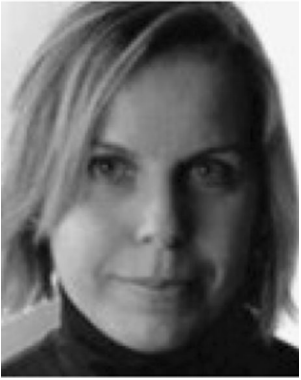

**Carmen Benavides** Ph.D. received her Bachelor degree in Industrial Technical Engineer from the University of León (Spain) in 1996 and her Master degree in Electronic Engineering from the University of Valladolid (Spain) in 1998. Carmen obtained her Ph.D. in Computer Science from the University of León in 2009 and she works as an Assistant Professor at the same University since 2001. Her research interests are focused on applied Knowledge Engineering techniques, practical applications of Software Defined Networks and Network Security. She has organized several congresses, and has presented and published different papers in Journals, Conferences and Symposia.

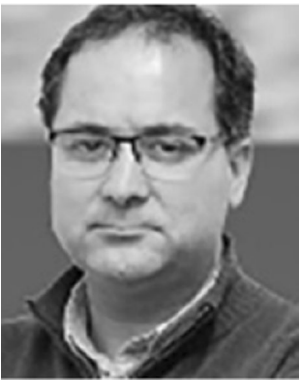

**Paulo Novais** Ph.D. degree in computer sciences and the Habilitation degree in computer science (Agregação ramo do conhecimento em Informática) from the University of Minho, Braga, Portugal, in 2003 and 2011, respectively, where he is currently a Full Professor of computer science with the Department of Informatics, School of Engineering, and a Researcher with the ALGORITMI Centre, in which he is the Leader of the Research Group ISlab–Synthetic Intelligence, and a Coordinator of the research line Computer Science and Technology. He is the Director of the Ph.D. Program in Informatics and also the Co-Founder and the Deputy Director of the master in law and informatics with the University of Minho. He started his career developing scientific research in the field of intelligent systems/artificial intelligence (AI), namely, in knowledge representation and reasoning, machine learning, and multi-agent systems. His interest, in the last years, was absorbed by different, yet closely related concepts of ambient intelligence, ambient assisted living, intelligent environments, behavioral analysis, conflict resolution, and the incorporation of AI methods and techniques in these fields. His main research aim is to make systems a little more smart, intelligent, and also reliable. He has led and participated in several research projects sponsored by Portuguese and European public and private institutions and has supervised several Ph.D. and M.Sc. students. He has co-authored over 300 book chapters, journal papers, and conference and workshop papers and books.

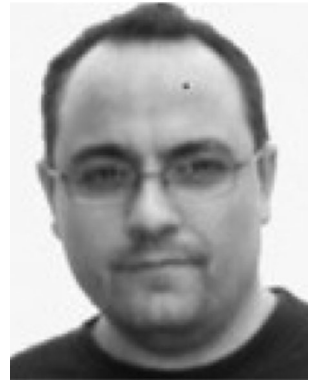

**Isaías García-Rodríguez** Ph.D. received his Bachelor degree in Industrial Technical Engineering from the University of León (Spain) in 1992 and her Master degree in Industrial Engineering from the University of Oviedo (Spain) in 1996. Isaías obtained his Ph.D. in Computer Science from the University of León in 2008, where he is currently a lecturer. His current research interests include practical applications of Software Defined Networks, Network Securityand applied Knowledge Engineering techniques. He has published different scientific papers in journals, Conferences and Symposia around the world.